\documentclass[3p,times]{elsarticle}

\usepackage{ecrc}
\usepackage{hyperref}
\usepackage{slashed}

\usepackage{epstopdf}

\volume{00}

\firstpage{1}

\journalname{Nuclear Physics A}

\runauth{N.~Armesto et al.}

\jid{nupha}

\jnltitlelogo{Nuclear Physics A}

\usepackage{graphicx}
\usepackage{amsmath,amssymb}

\usepackage{lineno}

\newcommand{\sqrtsNN}{\sqrt{s_{\rm \scriptscriptstyle NN}}}

\begin{document}

\begin{frontmatter}



\title{Heavy-ion physics studies for the Future Circular Collider}



\author[label1]{\normalsize N.~Armesto}
\author[label2]{A.~Dainese}
\ead{andrea.dainese@pd.infn.it}
\author[label3]{D.~d'Enterria}
\author[label4]{S.~Masciocchi}
\author[label5]{C.~Roland}
\author[label1]{\\C.A.~Salgado}
\author[label7]{M.~van Leeuwen}
\author[label3]{U.A. Wiedemann}

\address[label2]{INFN - Sezione di Padova, 35131 Padova, Italy}
\address[label1]{Departamento de F\'isica de Part\'iculas and IGFAE, Universidade de Santiago de Compostela, 15706 Santiago de Compostela, Galicia-Spain}
\address[label3]{Physics Department, CERN, CH-1211 Gen\'eve 23, Switzerland}
\address[label4]{EMMI and GSI Helmholtzzentrum fuer Schwerionenforschung GmbH, Darmstadt, Germany}
\address[label5]{Massachusetts Institute of Technology, Cambridge, MA 02139-4307, USA}
\address[label7]{Nikhef, National Institute for Subatomic Physics and Institute
for Subatomic Physics of Utrecht University, Utrecht, Netherlands}

\begin{abstract}
The Future Circular Collider (FCC) design study is aimed at assessing the physics potential and the technical feasibility
of a new collider
with centre-of-mass energies, in the hadron--hadron collision mode including proton and nucleus
beams, more than seven-times larger than the nominal LHC energies.
An electron--positron collider in the same tunnel is also considered as
an intermediate step, which would provide the electron--hadron option in the long term.
First ideas on the physics opportunities with heavy ions at the FCC are presented, 
 covering the physics of Quark--Gluon Plasma, gluon saturation, photon-induced collisions, as
well as connections with ultra-high-energy cosmic rays.
\end{abstract}


\end{frontmatter}



\section{Introduction}
\label{intro}

A five-year international design study called Future Circular Collider (FCC) has been launched by CERN in February 2014~\cite{FCCkickoff}. 
The main goal is a hadron collider with a centre-of-mass energy $\sqrt s$ of the order of 100~TeV for pp collisions in a new 80--100~km tunnel in the Geneva area. 
The target start of operation would be 2035--40.
Operating such machine with heavy ions is an option that is being considered in the accelerator design studies. 

For a centre-of-mass energy $\sqrt{s_{\rm pp}}= 100$~TeV for pp collisions, the relation $\sqrtsNN= \sqrt {s_{\rm pp}} \sqrt{Z_1 Z_2 / A_1 A_2}$ 
gives the energy in centre-of-mass per nucleon--nucleon collision of $\sqrtsNN = 39$~TeV for Pb--Pb ($Z=82$, $A=208$) and 63~TeV for p--Pb collisions. 
The rapidities of the proton and Pb beams are 11.6 and 10.6, respectively.
A first conservative estimate of the integrated luminosity for Pb--Pb collisions is of about 5~nb$^{-1}$ per month of running~\cite{schaumann}, that is about five times larger than
the current projection for the future LHC runs~\cite{rliup}.

The increased centre-of-mass energy and luminosity with respect to the LHC open new opportunities. Here, we present first general considerations on the following topics:
\begin{itemize}
\item high-density QCD and Quark--Gluon Plasma (QGP) in the final state of heavy-ion collisions (section~\ref{sec:QGP});
\item high-density QCD and gluon saturation in the initial state of heavy-ion collisions (section~\ref{sec:saturation});
\item photon-induced collisions and connections with cosmic-ray physics (section~\ref{sec:others}).
\end{itemize}
More details can be found in~\cite{DaineseKickoff}.

\vspace{-3mm}

\section{Study of the hot and dense Quark-Gluon Plasma in the final state of heavy-ion collisions}
\label{sec:QGP}

The QGP state produced in Pb--Pb collisions at $\sqrtsNN=39$~TeV is expected to have larger volume, lifetime, energy density and temperature 
than at the top LHC energy. These properties can be estimated by extrapolating the measurements of global event characteristics
at lower energies ---namely: the charged particle multiplicity, the transverse energy and the parameters extracted from Bose-Einstein (BE) correlations~\cite{extrap}.
To this purpose, the following parametrizations were used:
${\rm d}N_{\rm ch}/{\rm d}\eta$ at $\eta=0$ $\propto (\sqrtsNN)^{0.3}$;
total $N_{\rm ch} \propto 0.512\,  (\sqrtsNN)^{0.3} \ln s_{\rm \scriptscriptstyle NN} + 1.962$, with $\sqrtsNN$ in GeV;
${\rm d}E_{\rm T}/{\rm d}\eta$ at $\eta=0$ $\propto 0.46\,(\sqrtsNN)^{0.4}$;
BE homogeneity volume $\propto {\rm d}N_{\rm ch}/{\rm d}\eta$;  
BE decoupling time $\propto ({\rm d}N_{\rm ch}/{\rm d}\eta)^{1/3}$.  

\begin{figure}[!t]
\begin{center}
\includegraphics[width=0.41\textwidth]{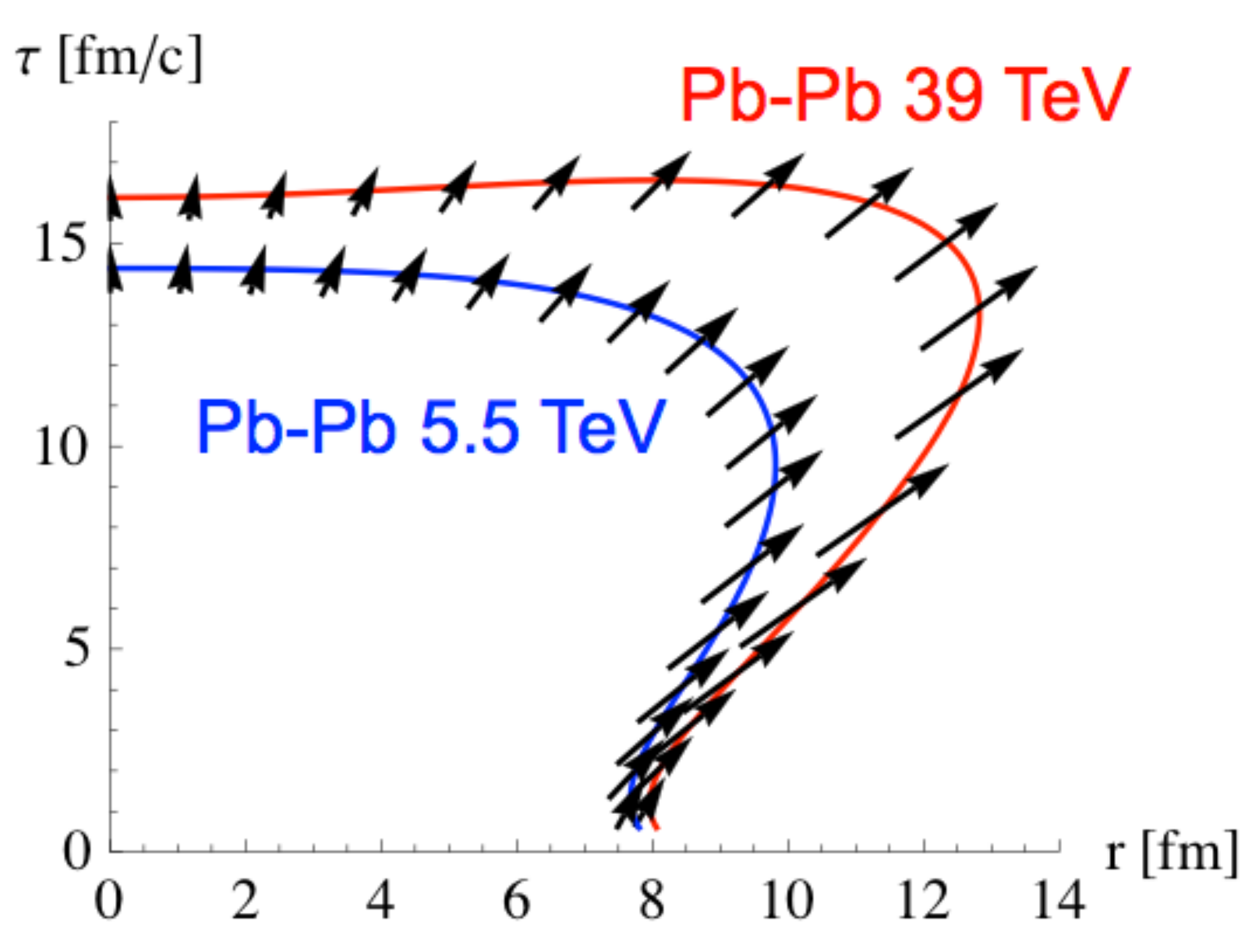}
\hfill
\includegraphics[width=0.38\textwidth]{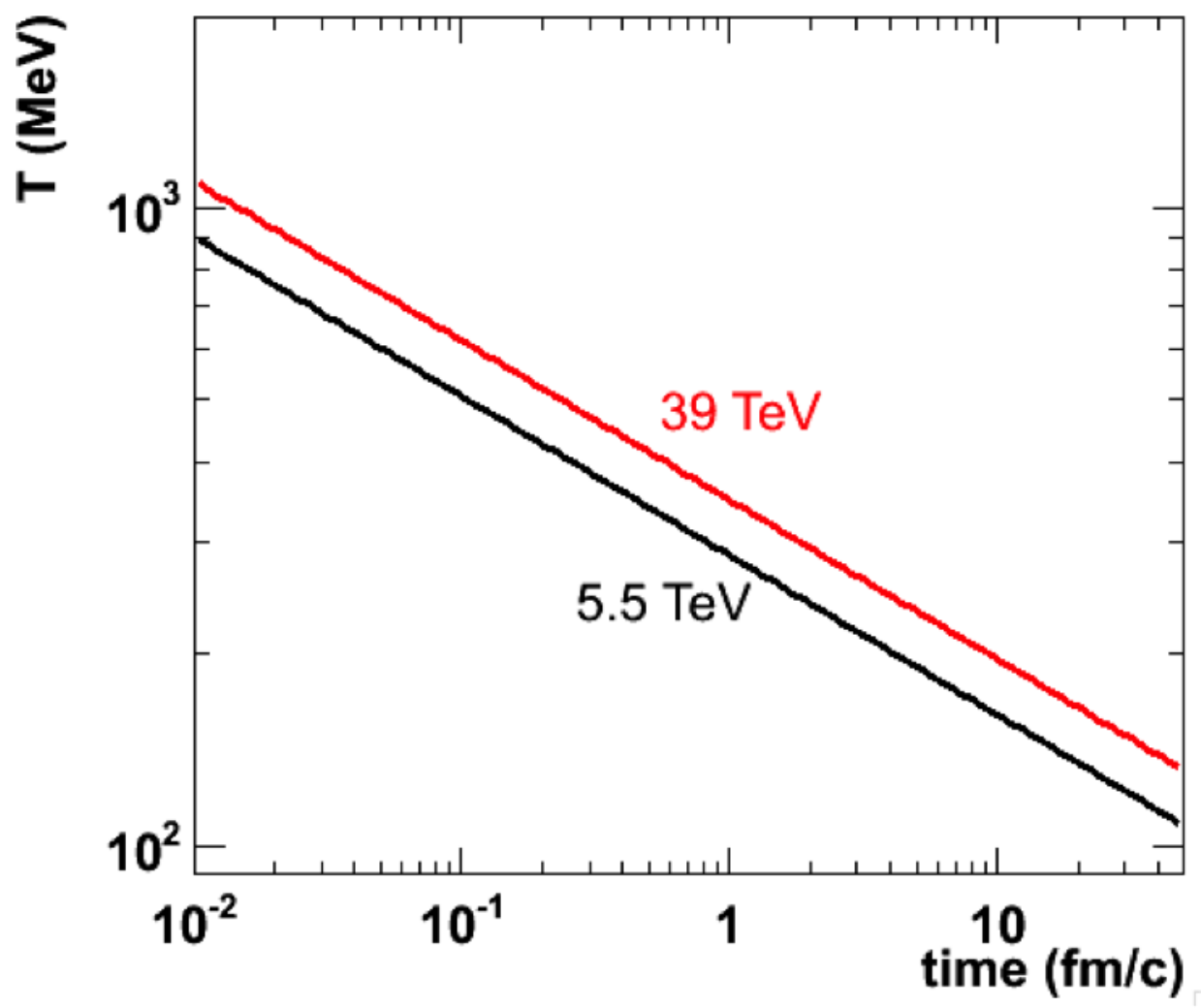}
\caption{Left: space-time profile of the freeze-out volume from a hydro-dynamic simulation of central Pb--Pb collisions at $\sqrtsNN=5.5$~TeV and 39~TeV~\cite{Florchinger}.
Right: time dependence of the QGP temperature,  obtained using the Stefan-Boltzmann relation between energy density and temperature and the Bjorken estimate of the energy density (see text for details).}
\label{fig:freezeout}
\end{center}
\end{figure}

The results are reported in Table~\ref{tab:PbPb} for central Pb--Pb collisions (0--5\%). The charged-particle multiplicity is expected to increase by 
about a factor of two from top LHC to FCC energy. The volume and lifetime of the system are expected to increase by a factor of two and by 20\%, 
respectively.  These factors are confirmed by a full hydro-dynamical simulation with input parameters shear-viscosity-over-entropy-density
ratio $\eta/s=1/4\pi$ and initial temperature such that the final ${\rm d}N_{\rm ch}/{\rm d}\eta$ is 3600~\cite{Florchinger}. 
The space-time profile of the freeze-out volume is shown in Fig.~\ref{fig:freezeout} (left). The length of the arrows
represents the strength of the radial flow field: a substantial increase can be seen.
The larger final volume and stronger flow field suggest that the collective effects would be enhanced. In addition, the two-fold larger multiplicity
may open the possibility to carry out flow measurements on an event-by-event basis.

\vspace{-1mm}

\begin{table}[h]
\caption{Global properties measured in central Pb--Pb collisions at $\sqrtsNN=2.76$~TeV and extrapolated to 5.5 and 39~TeV.
Note that the values measured at 2.76~TeV are reported without experimental uncertainties and are intended for comparison only.}
\small
\begin{center}
\begin{tabular}{lccc}
\hline
Quantity & Pb--Pb 2.76~TeV & Pb--Pb 5.5~TeV & Pb--Pb 39~TeV \\
\hline
${\rm d}N_{\rm ch}/{\rm d}\eta$ at $\eta=0$ & 1600 & 2000 & 3600 \\
Total $N_{\rm ch}$ & 17000 & 23000 & 50000 \\
${\rm d}E_{\rm T}/{\rm d}\eta$ at $\eta=0$ & 2~TeV & 2.6~TeV & 5.8~TeV \\
BE homogeneity volume $$ & 5000~fm$^3$  & 6200~fm$^3$ & 11000~fm$^3$ \\
BE decoupling time & 10~fm/$c$ &  11~fm/$c$ & 13~fm/$c$ \\
\hline
\end{tabular}
\end{center}
\label{tab:PbPb}
\end{table}


The Bjorken expression $\varepsilon(t) = \frac{1}{c\, t} \frac{1}{\pi R_A^2} {\rm d}E_{\rm T}/{\rm d}\eta$ estimates the time-dependent 
energy density of the system starting from the measured transverse energy.
Then, the Stefan-Boltzmann relation provides an estimate of the 
temperature evolution of the system: 
$T(t)= [\varepsilon(t)\,(30/\pi^2)/n_{\rm d.o.f.}]^{1/4}$, where $n_{\rm d.o.f.}=47.5$ is the number of degrees of freedom for a system with gluons and three quark flavours.
The energy density increases by a factor of two from LHC to FCC, reaching for example a value of 35~GeV/fm$^3$ at 1~fm/$c$. Figure~\ref{fig:freezeout} (right) shows
the time-dependence of the estimated temperature for top LHC and for FCC energy.  The increase, at given time, is modest, but the thermalization time of the system 
(initial time of the QGP evolution) can be expected to be smaller at FCC than at LHC, where it is usually taken as $\tau_0=0.1$~fm/$c$. If the thermalization time 
is significantly lower than 0.1~fm/$c$, the initial temperature could be as large as $T_0\approx800$~MeV. 

\begin{figure}[!t]
\begin{center}
\includegraphics[width=0.35\textwidth,height=4.5cm]{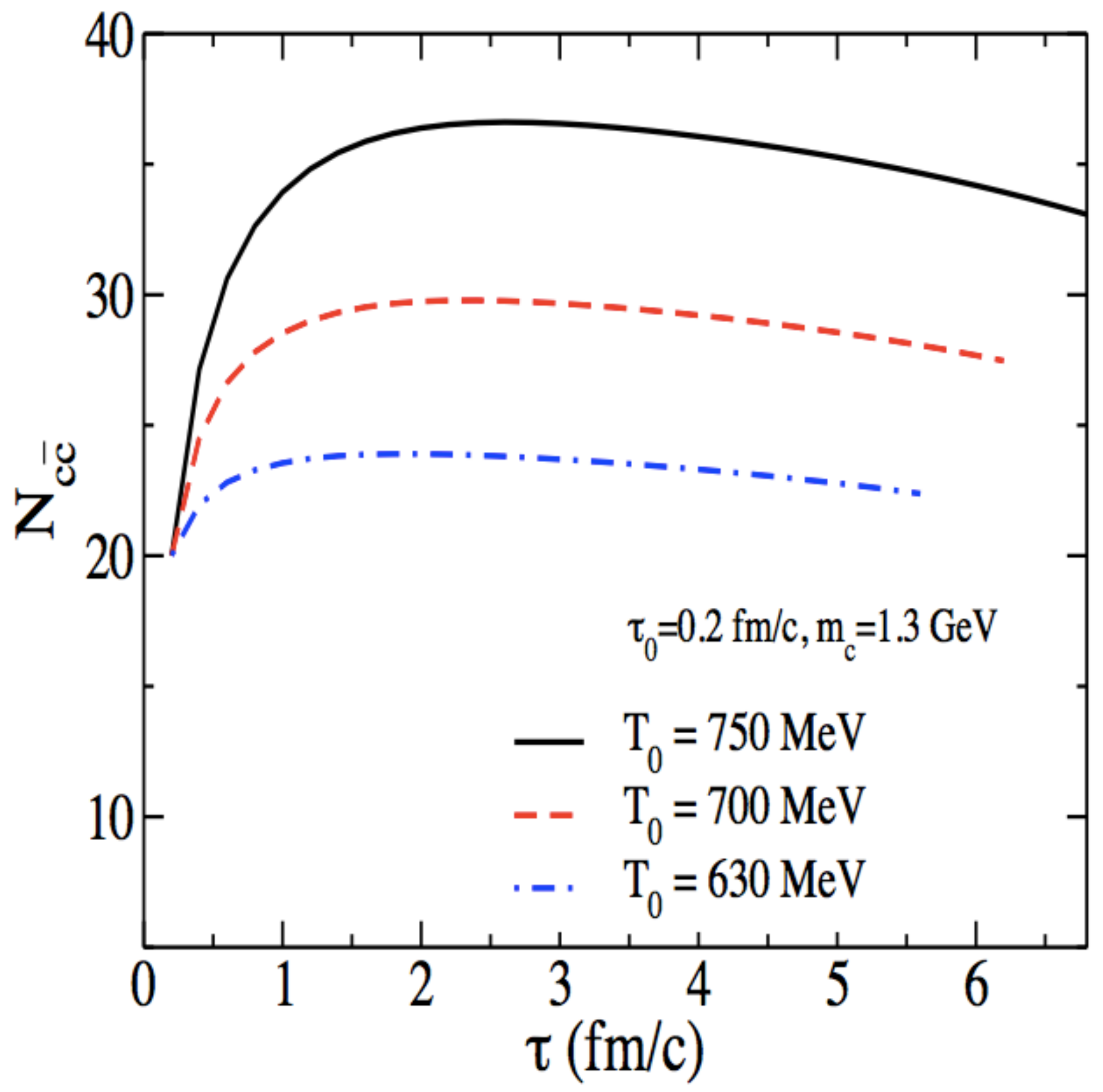}
\hfill
\includegraphics[width=0.40\textwidth]{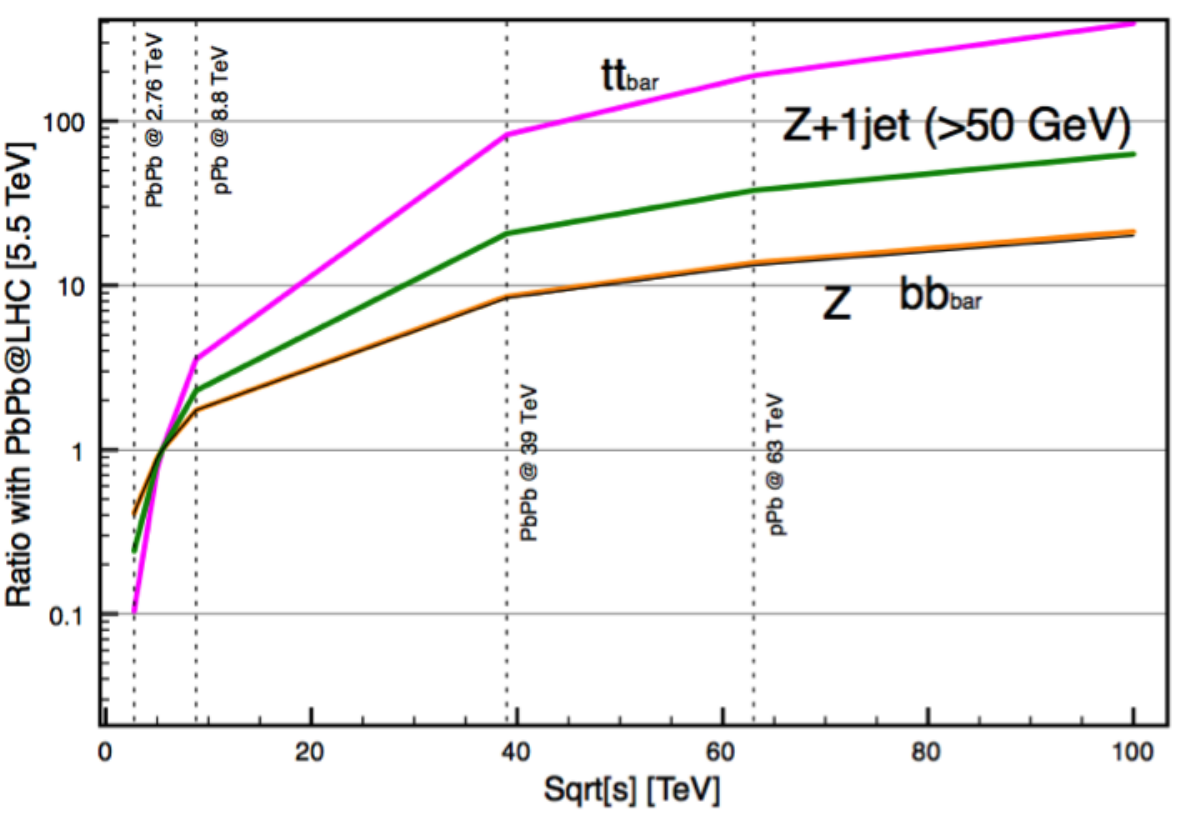}
\caption{Left: time-evolution of the charm-anticharm yield (per unit of rapidity at central rapidity) for central Pb--Pb collisions 
at $\sqrtsNN=5.5$~TeV~\cite{ko}.
Right: hard scattering cross sections for various processes as a function of centre-of-mass energy, obtained at NLO with the MCFM code~\cite{mcfm}, normalised to their value at 5.5~TeV.}
\label{fig:charm}
\end{center}
\end{figure}

An interesting consequence of the increase in the temperature of the system could be a sizeable production of secondary charm-anticharm ($\rm c\overline c$) pairs from partonic
interactions during the hydrodynamical evolution of the system.
Calculations for top LHC energy indicate that this secondary production a) can become of the same order of the initial production in hard scattering processes
and b) it is very sensitive to the initial temperature and temperature evolution of the QGP~\cite{uphoff,ko}. As an example, Fig.~\ref{fig:charm} (left) shows the time evolution of the 
number of $\rm c\overline c$ pairs per unit of rapidity at central rapidity (the value at $t=0$ represents the production from hard scattering). The secondary charm production would yield an enhancement of charmed hadron 
production in the very-low-$p_{\rm T}$ region, with respect to the expectation from binary scaling of the production in pp collisions. This enhancement potentially provides a handle on the
temperature of the QGP. In addition, the abundance of charm quarks in the QGP is expected to have an effect on its equation of the state: lattice QCD calculations show a sizeable 
increase of $P/T^4$ (which is proportional to the number of degrees of freedom) when the charm quark is included and the system temperature is larger than about 400~MeV~\cite{ratti}.

The larger energy and luminosities will make new, rarer, hard probes available. 
Figure~\ref{fig:charm} (right) shows the energy dependence of a selection of hard cross sections computed at NLO accuracy with the MCFM code~\cite{mcfm}, 
normalised to their value at 5.5~TeV: the cross sections for 
top, Z$^0$+$1\,{\rm Jet}\,(p_{\rm T}>50~{\rm GeV}/c)$, bottom and Z$^0$ increase by factors 80, 20, 8 and 7, respectively. 
In the following, we give an example of the very large statistics that could be obtained for currently unexplored observables, like top pair production.
Starting from the projections made by the CMS Collaboration for future LHC runs~\cite{CMStop}, we have estimated that an experiment at FCC could record about $2.5\cdot 10^4$ events with the topology  ${\rm t\overline t\to b\overline b }+ \ell^+\ell^- + E_{\rm T}^{\rm missing}$ 
fully reconstructed using a Pb--Pb sample with integrated luminosity of $5~{\rm nb^{-1}}$.

\vspace{-3mm}

\section{Study of the saturation of the gluon density in the initial state of heavy-ion collisions}
\label{sec:saturation}

\begin{figure}[!t]
\begin{center}
\includegraphics[width=0.37\textwidth]{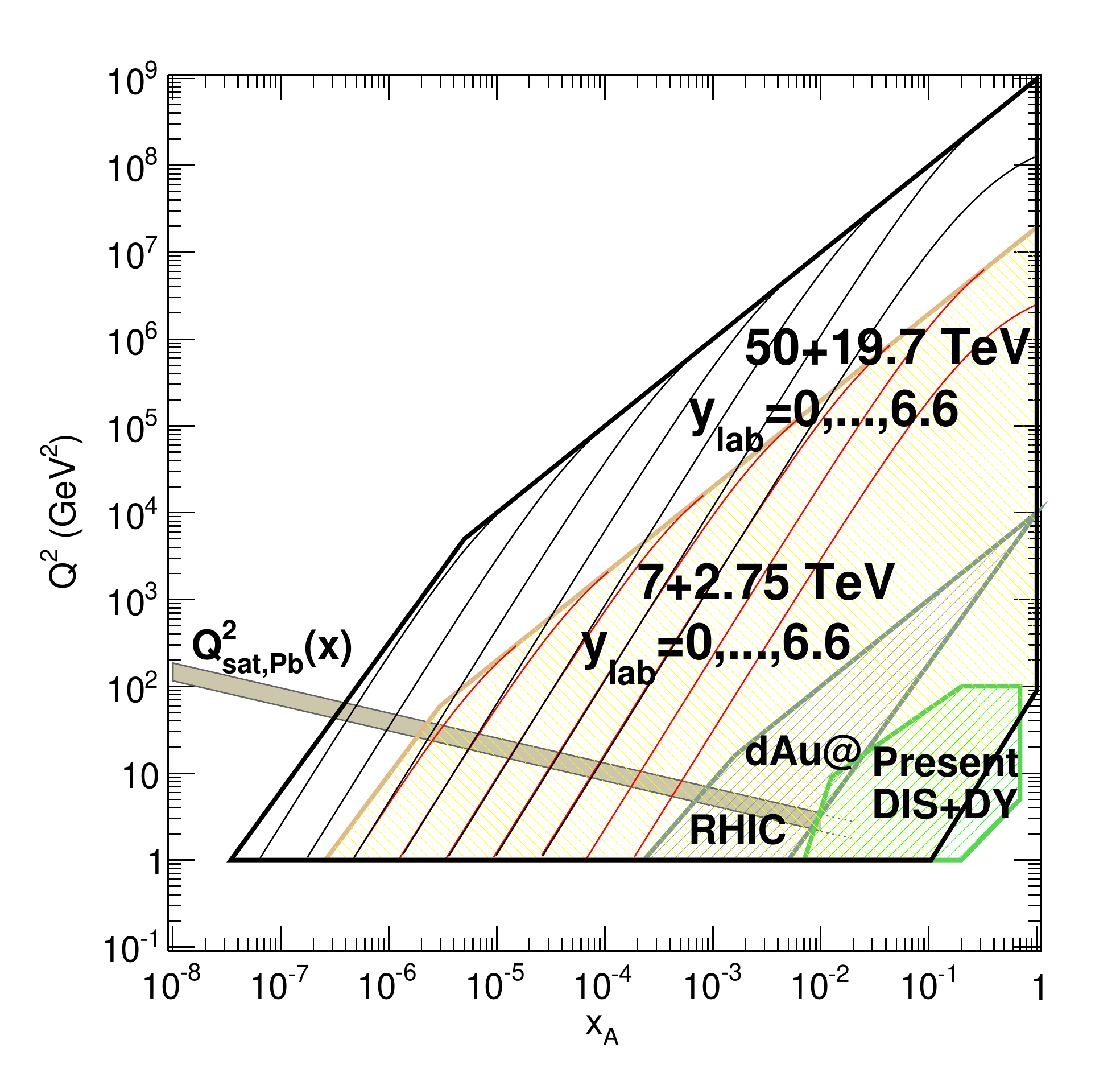}
\hfill
\includegraphics[width=0.40\textwidth]{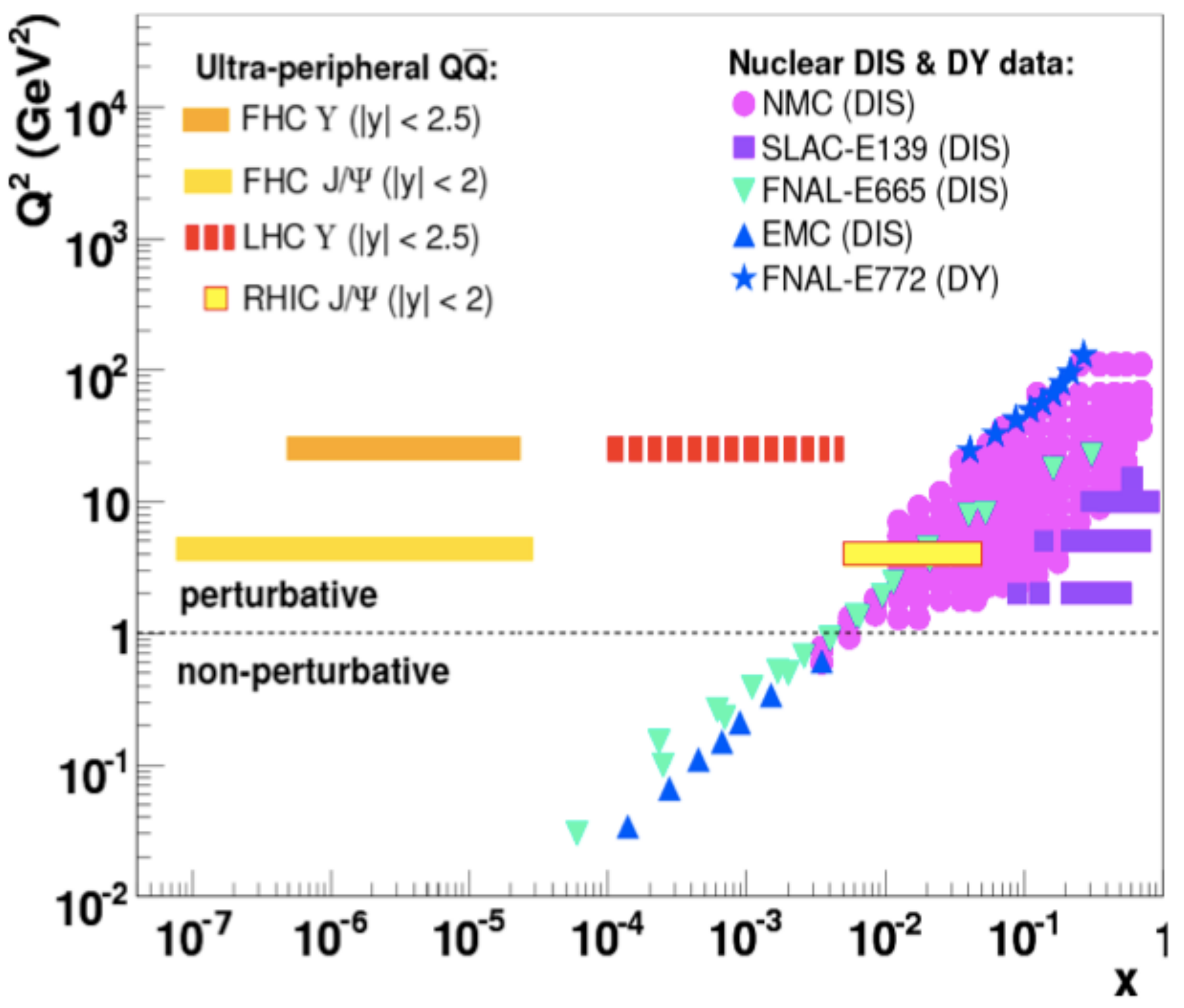}
\caption{Kinematic coverage in the $(x,Q^2)$ plane. Left: coverage for p--Pb collisions at RHIC, LHC and FCC energies, computed as  $x(y,Q^2) =  Q\cdot {\rm e}^{-y} / \sqrtsNN$;
Constant-rapidity lines and an estimation of the saturation scale for Pb nuclei are shown.
 Right: coverage 
for quarkonium production in ultra-peripheral collisions at LHC and FCC energies (horizontal bars), compared with existing data on nuclear DIS and Drell-Yan.}
\label{fig:smallx}
\end{center}
\end{figure}

Proton--nucleus, nucleus--nucleus and electron--nucleus collisions at very high energy provide a unique opportunity to study the dynamics of high-density gluon densities in the initial 
nuclear state. The increase of the gluon density towards low virtuality $Q^2$ and low momentum fraction $x$ is expected to be at some point tamed by 
a saturation effect (see e.g.\,\cite{Albacete:2014fwa} for a recent review). 
When density becomes large, the dilute linear evolution of parton densities must break down and non-linear terms have to be considered.
Such non-linear 
terms describe merging processes of the type $gg\to g$ that balance the growth of the gluon density given by the splitting processes of the type $g\to gg$.
Saturation effects become sizeable in processes with virtualities smaller than a few times the saturation scale $Q^2_{\rm sat}$,
which is estimated to scale as $Q^2_{\rm sat}\sim A^{1/3}/x^{1/3}\sim A^{1/3}(\sqrt s\, {\rm e}^{+y})^{1/3}$ ($A$ is the mass number of the nucleus). 
The saturation scale can be increased (thus making its effects more prominent) by increasing $\sqrt s$, by using large nuclei instead of protons and by 
measuring at large rapidity $y$. 
Figure~\ref{fig:smallx} (left) shows the kinematic coverage of the $(x,Q^2)$ plane with p--Pb collisions at top LHC ($\sqrtsNN=8.8$~TeV) and FCC (63~TeV) energies. 
The constant-rapidity lines from 0 (right) to 6.6 (left) and
an estimation of the $x$-dependence of the saturation scale for Pb nuclei are also shown.
The FCC extends the coverage by almost one order of magnitude at low $x$, down to $\sim 10^{-7}$. The access is extended down to $10^{-6}$ even in the region $Q^2>10~$GeV$^2$, 
which can be explored with probes, like heavy quarks and charmonium, that are perturbatively calculable, thus under more robust theoretical control.

\vspace{-3mm}

\section{Photon-induced collisions and connection with cosmic-ray physics}
\label{sec:others}

Charged hadrons accelerated at very high energies generate strong electromagnetic fields,
equivalent to a flux of quasi-real photons, which can be used to study high-energy $\gamma$--$\gamma$,
$\gamma$--p and $\gamma$--A processes in ultra-peripheral collisions (UPCs) where the colliding systems
pass close to each other without interacting hadronically. As shown in the right panel of Fig.~\ref{fig:smallx}, 
the small-$x$ coverage would be strongly enhanced at FCC energies
also in the sector of ultra-peripheral Pb--Pb and p--Pb collisions. In this case, quarkonium production 
probes the gluon densities at $x$ values down to $10^{-7}$, that is more than two orders of magnitude lower than at LHC energy.

Cosmic-ray showers are initiated by proton--nucleus and nucleus--nucleus collisions in the upper atmosphere (mainly p--N and Fe--N). 
High-energy nucleus--nucleus collisions provide important information for the understanding of these primary interactions and 
for the
tuning of the hadronic Monte Carlo generators used to determine the energy and identity of the
incoming cosmic ray~\cite{uhecr}.
Pb--Pb collisions at the LHC are equivalent to primary cosmic-ray interactions with incoming nucleus energy of a few $10^{16}$~eV, which is about three orders
of magnitude lower than that of the highest observed cosmic rays energies. Pb--Pb collisions at the FCC would correspond to 
a primary cosmic-ray interaction energy of a few $10^{18}$~eV, thus reducing to only one order of magnitude the 
extrapolation to the highest energies observed on Earth at the so-called GZK cutoff.

\end{document}